\documentclass{aip-cp}

\usepackage[numbers]{natbib}
\usepackage{rotating}
\usepackage{graphicx}

\begin{document}

\title{Helicity Asymmetry Measurement for $\pi^0$ Photoproduction 
	on the CLAS Frozen Spin Target}

\author[aff1]{Diane Schott}
\author[aff1]{William Briscoe}
\author[aff1]{Igor Strakovsky\corref{cor1}}

\affil[aff1]{Department of Physics, The George Washington 
	University, Washington, D.C. 20052, USA}
\corresp[cor1]{Corresponding author: igor@gwu.edu}
\maketitle

\centerline{for CLAS Collaboration}
\vspace{0.2in}

\begin{abstract}
The measurement was performed with circularly polarized photons 
incident on longitudinally polarized target in Hall~B at the 
Thomas Jefferson National Accelerator Facility. The helicity 
asymmetry $E$ for $\vec{\gamma}\vec{p}\to\pi^0p$ was determined 
at CM energies between 1450~MeV and 2050~MeV and compared to 
the SAID, MAID, and BnGa partial-wave analyses.
\end{abstract}

\section{Intoduction}

The second generation of CEBAF Large Acceptance Spectrometer 
(CLAS) photoproduction experiments used the FROzen Spin 
Target (FROST), which allowed double-polarization measurements 
on proton.  With the FROST, we measure all beam-target 
double-polarization observables for single pseudoscalar meson 
photoproduction and for two charged pion production as well. 
This makes possible the complete experiment, which measures 
enough observables for unambiguous and direct reconstruction 
of the reaction amplitude~\cite{Andy}. CLAS collected data 
with linearly and circularly polarized photons and 
longitudinally and transverse polarized target. Many of the 
observables in this experiment were measured for the first 
time.

The entire data set is invaluable for a multi-channel analysis. 
The value of these data is more than just its broad coverage 
for different reaction channels and observables. The real 
strength of this program is its measurement of everything 
under the same controlled conditions with the same systematic 
uncertainty.  This provides much stronger constraints for 
subsequent analyses of the properties of contributing nucleon
resonances. With this greater understanding of these 
observables, effects of higher spin resonances can be 
investigated~\cite{Saghai}.

In this contribution, we present a measurement of the 
double-polarization observable $E$ in the $\vec{\gamma}
\vec{p}\to\pi^0p$ reaction of circularly polarized 
photons with longitudinally polarized protons
\cite{proposal}. The full energy coverage is E$_\gamma$ 
= 466 -- 1825~MeV (W = 1325 -- 2075~MeV).

\section{Experiment}

The experiment was performed at the Thomas Jefferson 
National Accelerator Facility (JLab). Data were taken 
within CLAS $G9A$ run group, November 2007 through 
February 2008. Longitudinally polarized electrons from 
the CEBAF accelerator with energies of $E_e$ = 1.465~GeV 
and 2.478~GeV were incident on the thin bremsstrahlung
radiator of the Hall-B Photon Tagger~\cite{Sober} and 
producing circularly polarized tagged photons in the 
energy range between $E_\gamma$ = 466 -- 1825~MeV.

The degree of circular polarization of the photon 
beam, $P_\odot$, depends on the ratio $x = 
E_\gamma/E_e$ and increases from zero to the 
degree of incident electron-beam polarization, $P_e$, 
monotonically with photon energy~\cite{Max}
\begin{equation}
        P_\odot = P_e~\frac{4x - x^2}{4 - 4x + 3x^2} .
\label{eq:eq3}
\end{equation}
Measurements of the electron-beam polarization were 
made routinely with the Hall-B Moeller polarimeter. 
The average value of the electron-beam polarization 
was found to be $P_e = 0.84\pm 0.04$. The 
electron-beam helicity was pseudo-randomly flipped 
between $+1$ and $-1$ with a 30~Hz flip rate.

The collimated photon beam irradiated the FROST
\cite{FROST} at the center of the CLAS~\cite{CLAS}. 
Frozen beads of butanol ($C_4H_9OH$) inside a 50~mm 
long target cup were used as target material. The 
protons of the hydrogen atoms in this material were 
dynamically polarized along the photon-beam 
direction and polarization was frozen. The degree of 
polarization on average was $P_z = 0.82\pm 0.05$. The 
proton polarization was routinely changed from being 
aligned along the beam axis to being anti-aligned. 
Quasi-free photoproduction off the unpolarized, bound 
protons in the butanol target constituted a background. 
Data were taken simultaneously from an additional 
carbon target down-stream of the butanol target to 
allow for the determination of the bound-nucleon 
background. 

\begin{figure}[htb!]
\centerline{
        \includegraphics[width=3.3in, height=1.4in, angle=0]{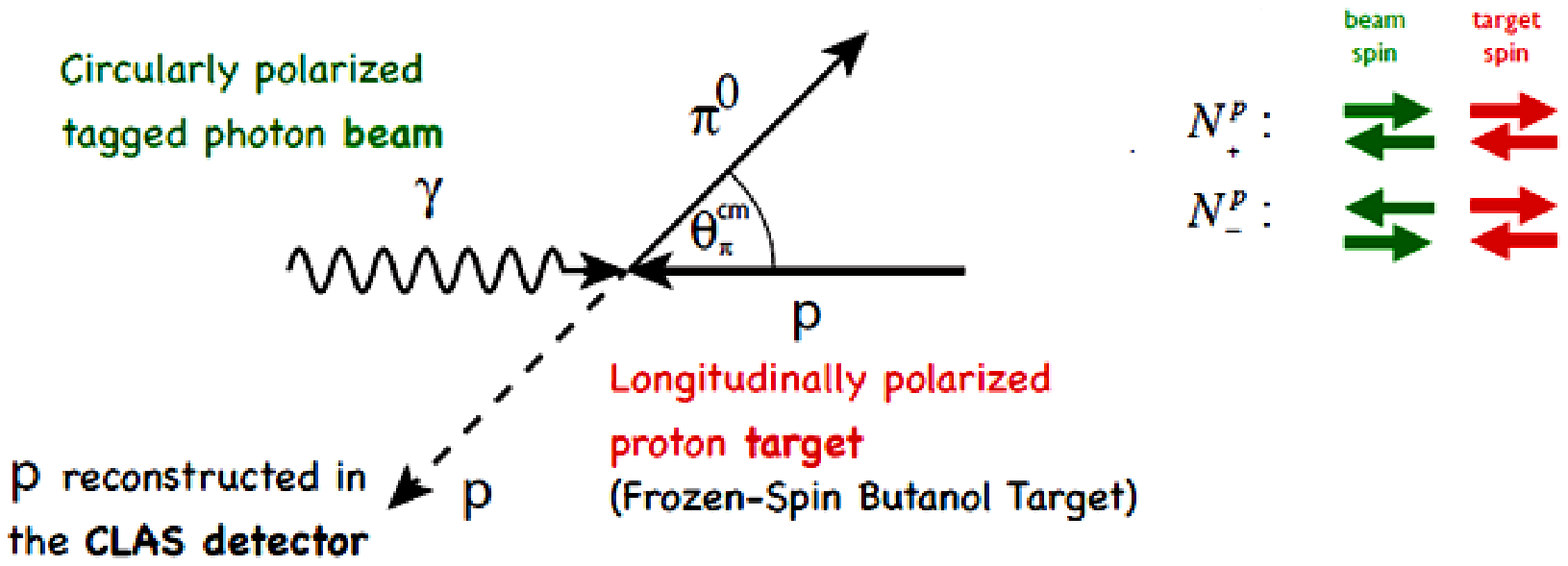}}

  \caption{Kinematics for $\vec{\gamma}\vec{p}\to\pi^0p$.} 
	\label{fig:fig1}
\end{figure}

The missing mass technique $\gamma p \to pX$, where 
$X$ is $\pi^0$ was used to identify the reaction. 
Final-state protons were detected in CLAS. The 
particle detectors used in this experiment were a 
set of plastic scintillation counters close to the 
target to measure event start times (Start Counter)
\cite{Sharab}, drift chambers~\cite{Mac} to 
determine charged-particle trajectories in the 
magnetic field within CLAS, and scintillation 
counters for flight-time measurements~\cite{Elton}. 
Coincident signals from the photon tagger, start-, 
and time-of-flight counters constituted the event 
trigger. Data from this experiment were taken in 
seven groups of runs with various electron-beam 
energies and beam/target polarization orientations. 
Events with one and only one positively charged 
particle and no other charged particles detected 
in CLAS were considered.  The protons were 
identified by their charge (from the curvature of 
the particle track) and by using the 
time-of-flight technique.  Kinematics of 
$\vec{\gamma}\vec{p}\to\pi^0p$ is given in 
Fig.~\ref{fig:fig1}.
The polarized cross section is in $\vec{\gamma}\vec{p}\to
\pi^0p$ case given by~\cite{Sandy}.
\begin{equation}
        \frac{d\sigma}{d\Omega} = \frac{d\sigma}
       {d\Omega}_0~(1 - P_z~P_\odot~E) ,
\label{eq:eq1}
\end{equation}
where $\frac{d\sigma}{d\Omega}_0$ is the unpolarized 
cross section, $P_z$ and $P_\odot$ are the target and 
beam polarizations, respectively. The observable $E$ 
is the helicity asymmetry of the cross section,
\begin{equation}
        E = \frac{d\sigma_{1/2} - d\sigma_{3/2}}
       {d\sigma_{1/2} + d\sigma_{3/2}}
\label{eq:eq2}
\end{equation}
for aligned, total helicity $h=3/2$ and anti-aligned,
$h=1/2$ photon and proton spins. These data are fitted
using three different PWA approaches - SAID~\cite{CM12},
MAID~\cite{MAID}, and BnGa~\cite{BnGa}. The resulting
consistency of helicity amplitudes for the dominant
resonances demonstrates that the PWA results are
largely driven by the data alone; the modest
differences gauge the model-dependence. This
consistency provides an excellent starting point to
search for new resonances (``missed resonance" 
problem).

\section{Experimental Data}

The asymmetry $E$ was determined in 256 kinematic bins 
of $W$ ($\Delta W = 50~MeV$) and $\cos\theta$
($\Delta(\cos\theta) = 0.1$), where $W$ is the CM energy 
and $\theta$ is the pion CM angle with respect to the 
incident photon momentum direction.

\begin{figure}[htb!]
\centerline{
	\includegraphics[width=3.1in, height=2.8in, angle=0]{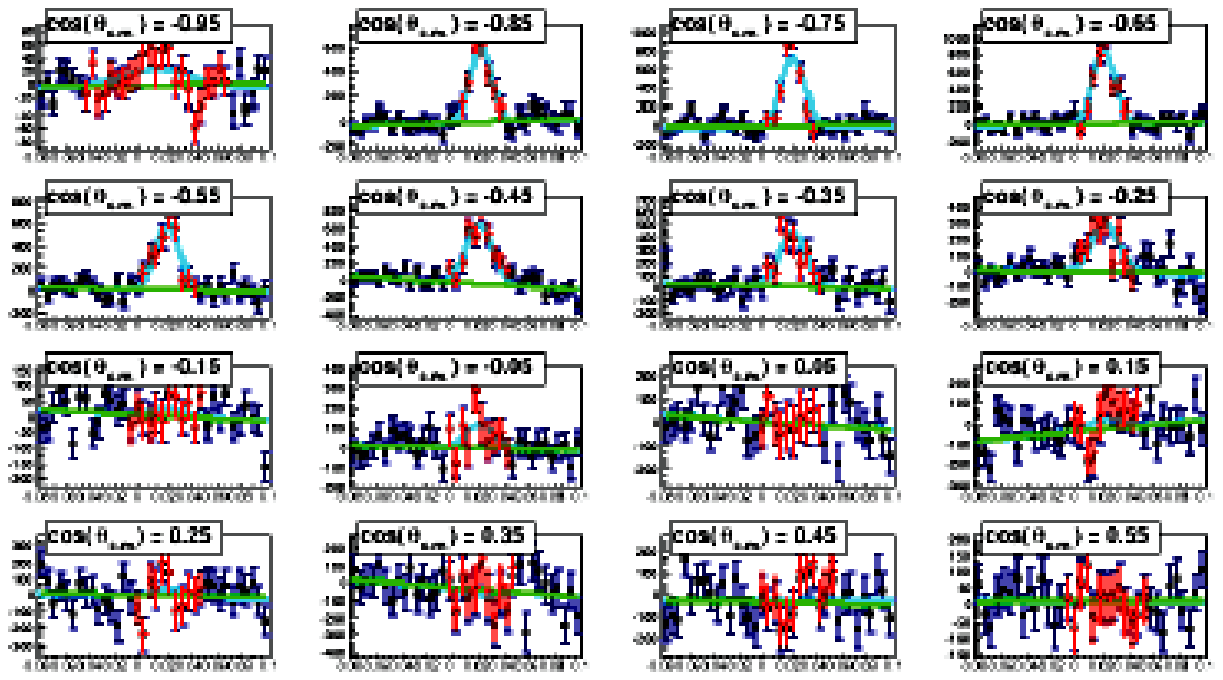}
	\includegraphics[width=3.1in, height=2.8in, angle=0]{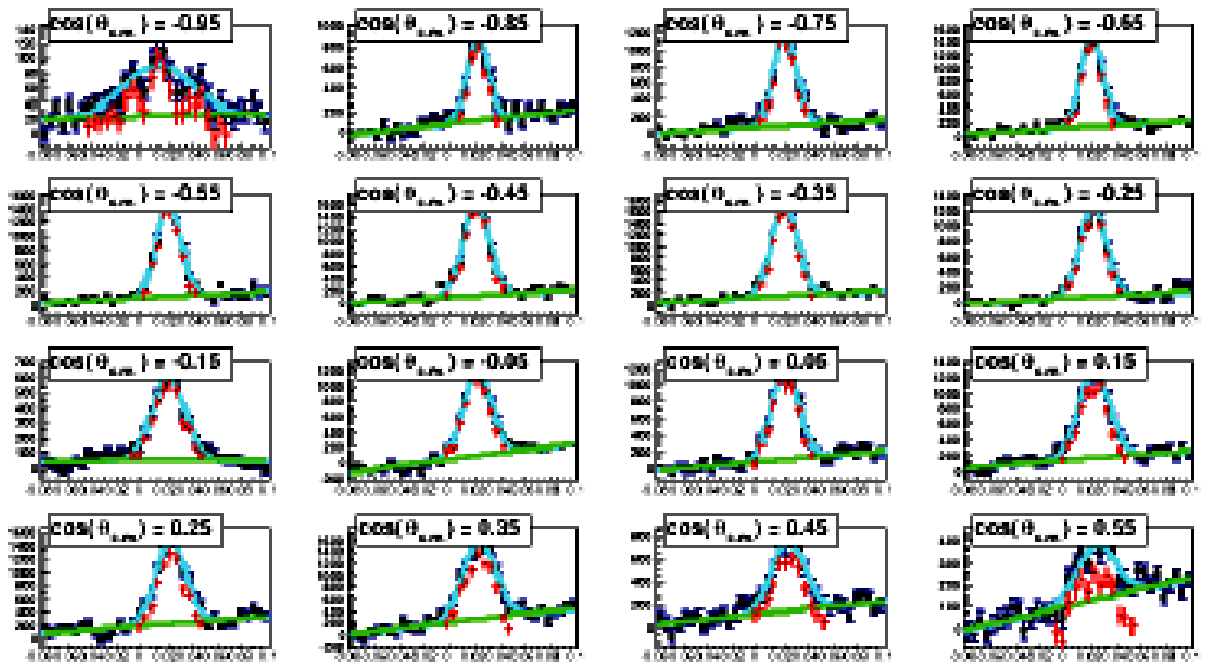}}

  \caption{Sample for yield determination at W = 1475~MeV.
	Numerator (left) and denominator (right) for 
	Eq.~(\protect\ref{eq:eq4}).} 
	\label{fig:fig2}
\end{figure}

The asymmetry $E$ was extracted from the polarized 
yields, $N^p_+$ and $N^p_-$ (Fig.~\ref{fig:fig1}), of 
$\vec{\gamma}\vec{p}\to\pi^0p$ events for total 
helicities $h = 3/2$ and $h = 1/2$, respectively, and 
the average beam and target polarizations,
\begin{equation}
        E = -\frac{1}{P_z~P_\odot}~\frac{N^p_+ - N^p_-}
	{N^p_+ - N^p_-} . \label{eq:eq4}
\end{equation}
Yields were determined using Gaussian plus polynomial 
to fit peak within $2~\sigma$ (Fig.~\ref{fig:fig2}).

\begin{figure}[htb!]
\centerline{
	\includegraphics[width=2.5in, height=2.1in, angle=0]{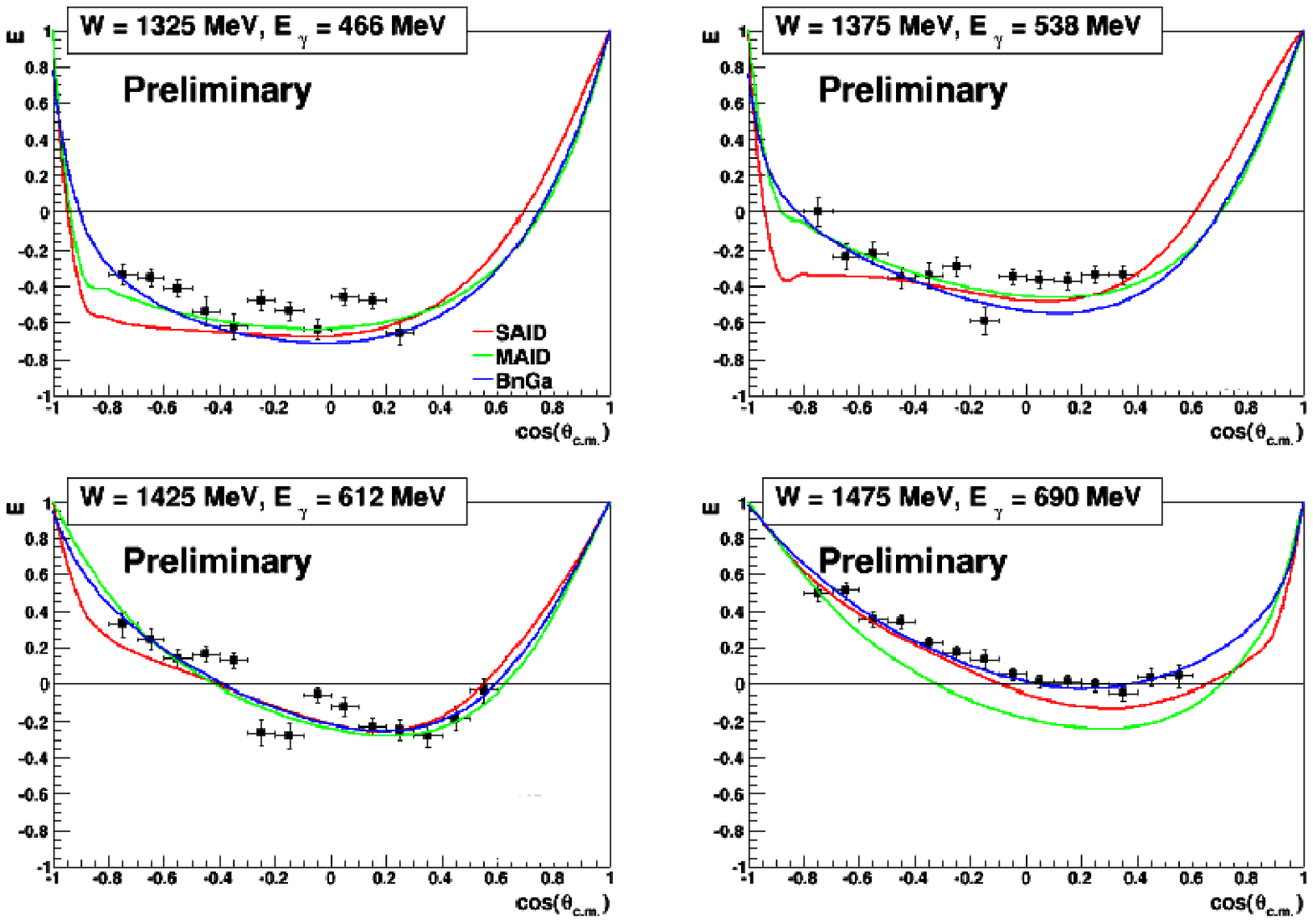}~~~~~~
	\includegraphics[width=2.5in, height=2.1in, angle=0]{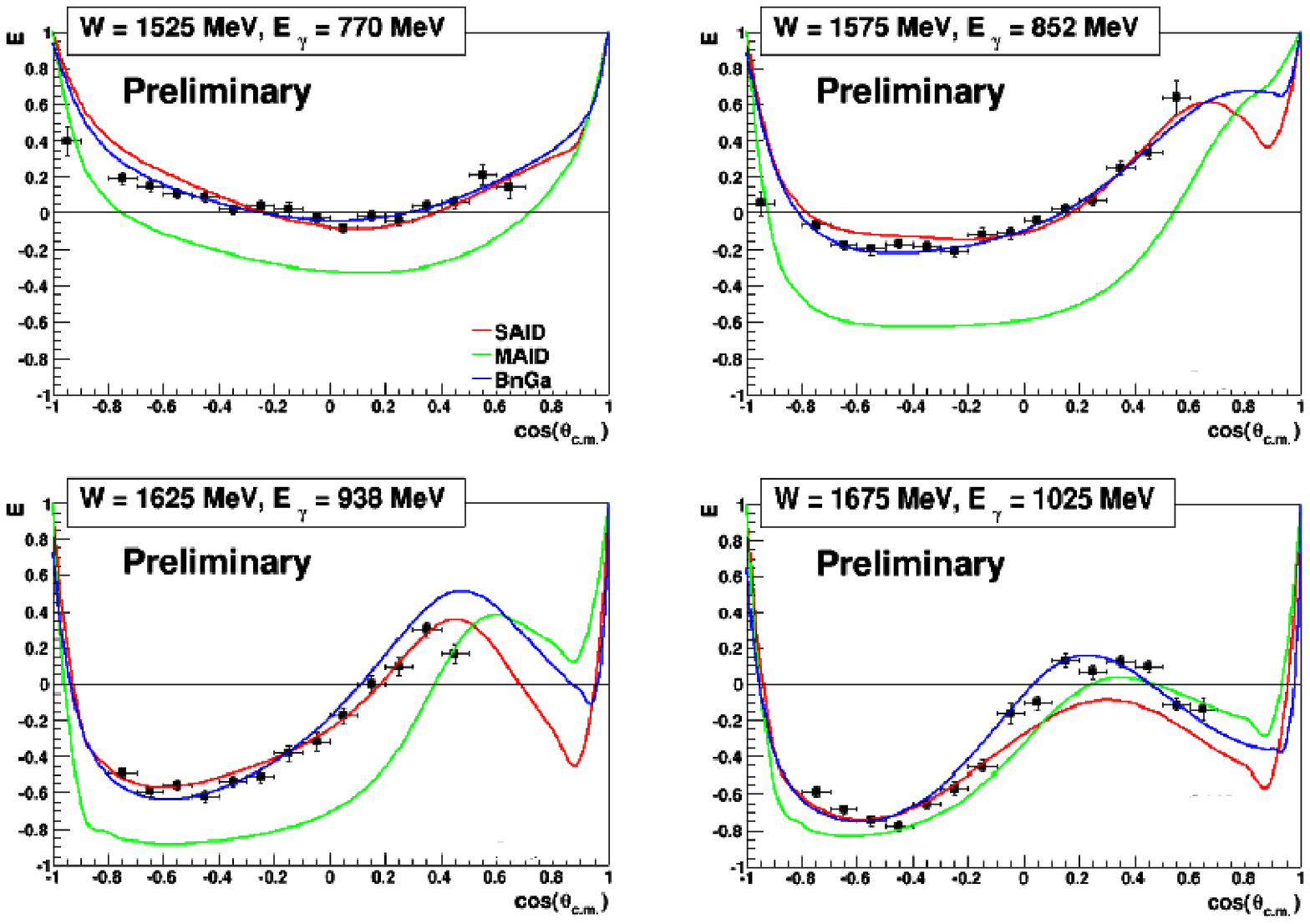}}

  \caption{Preliminary double polarization asymmetry $E$ 
	for $\vec{\gamma}\vec{p}\to\pi^0p$ at E$_\gamma$ 
	= 466 -- 1025~MeV versus pion $\cos$ CM production 
	angle. Photon energy is indicated by asymmetry $E$, 
	while the CM total energy is indicated by $W$.  
	Red solid (green solid) lines correspond to the 
	SAID CM12~\protect\cite{CM12} (MAID07
	\protect\cite{MAID}) predictions. Black solid 
	lines give the BG2011-02 BnGa~\protect\cite{BnGa} 
	predictions.} \label{fig:fig3}
\end{figure}

The preliminary results for asymmetry $E$ were compared with PWA 
predictions from SAID, MAID and BnGa groups (Figs.~\ref{fig:fig3}). 
They are in agreement at low energies but start to deviate at 
higher energies.
\begin{figure}[htb!]
\centerline{
        \includegraphics[width=2.5in, height=2.1in, angle=0]{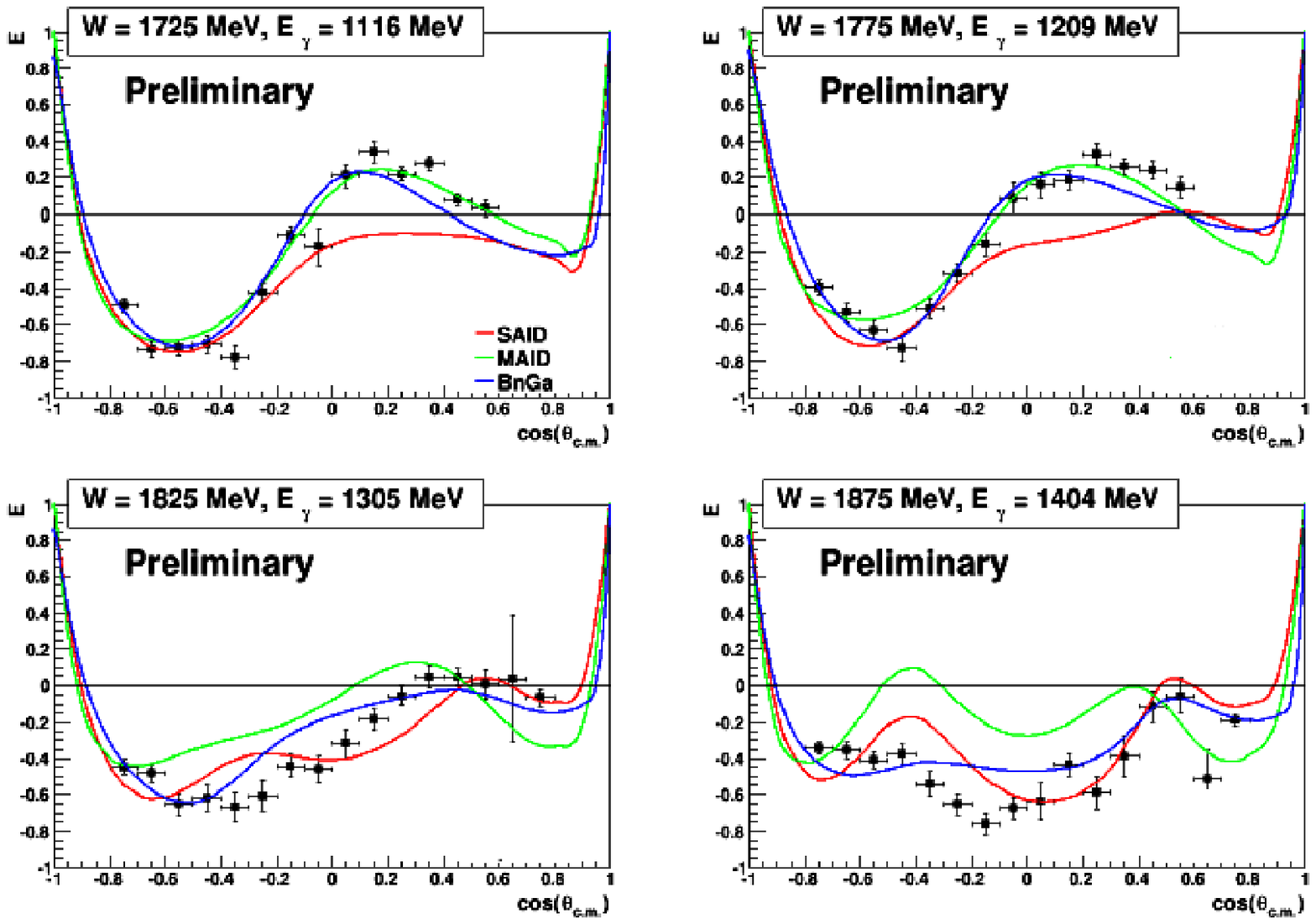}~~~~~~
        \includegraphics[width=2.5in, height=2.1in, angle=0]{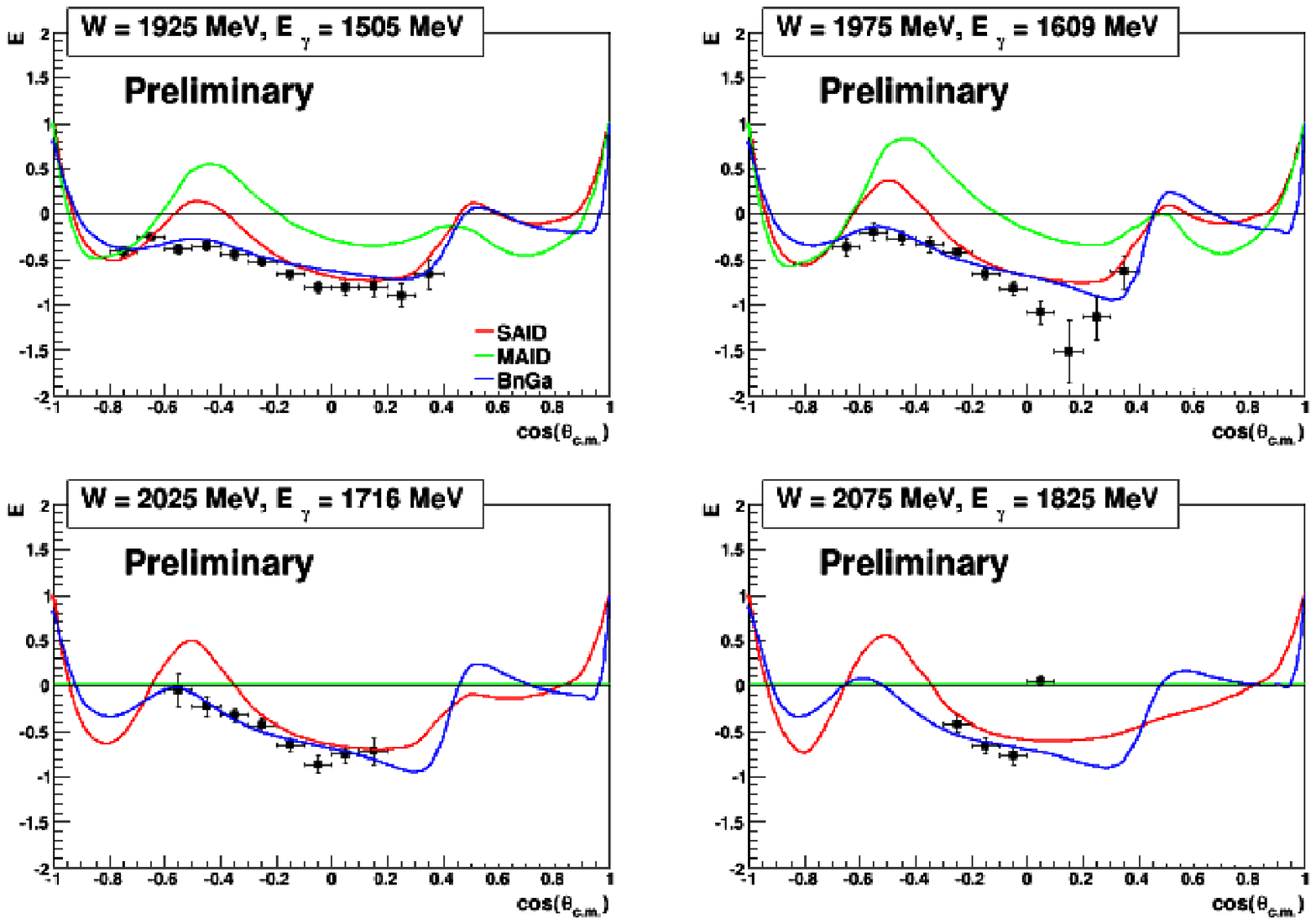}}

  \caption{Preliminary double polarization asymmetry $E$ 
	for $\vec{\gamma}\vec{p}\to\pi^0p$ at E$_\gamma$ 
	= 1116 -- 1825~MeV versus pion $\cos$ CM production 
	angle. Notation as in Fig.~\protect\ref{fig:fig3}.} 
	\label{fig:fig4}
\end{figure}

\begin{figure}[htb!]
\centerline{
       \includegraphics[width=3in, height=4in, angle=90]{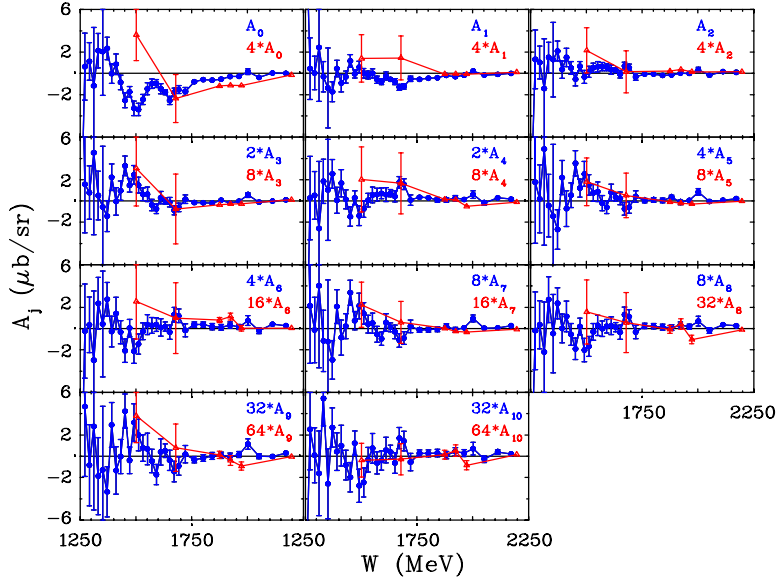}}

        \caption{Coefficients obtained from fitting the
                recent asymmetry $E$ for $\vec{\gamma}\vec{p}
		\to\pi^+n$, came from CLAS~\protect\cite{Steffen}
                (blue filled circles) and $\vec{\gamma}\vec{p}
                \to\pi^0p$, came from
                CBELSA~\protect\cite{CBELSA} (red filled
                triangles) with Legendre polynomials. The
                error bars of all values represent A$_j$
                uncertainties from the fits in which the
                statistical uncertainties were only used.}
                \label{fig:fig5}
\end{figure}

Beyond the SAID PWA, we plan the Legendre analysis for CLAS 
$E$ measurements for both $\vec{\gamma}\vec{p}\to\pi^+n$
\cite{Steffen} and new $\vec{\gamma}\vec{p}\to\pi^0p$ as we
did recently for the CLAS $\vec{\gamma}\vec{p}\to\pi^+n$
and $\vec{\gamma}\vec{p}\to\pi^0p$ $\Sigma$ asymmetry
measurements~\cite{Mike}. Unfortunately, recent CBELSA 
asymmetry $E$ for $\vec{\gamma}\vec{p}\to\pi^0p$~\cite{CBELSA} 
is insufficient for that study because of so broad energy 
binning ($\Delta W = 300 - 500~MeV$) (Fig.~\ref{fig:fig5}).

\section{Conclusion}

In summary, we have presented measurements of the 
double-polarization observable $E$ in the $\vec{\gamma}
\vec{p}\to\pi^0p$ up to W = 2075~MeV over a broad 
angular range. The data for $E$ asymmetry is the 
part of the FROST program at JLab. Results are 
consistent with PWA predictions (SAID, MAID, and 
BnGa PWA groups) at lower energies and will offer 
more results to be fit in the higher energy ranges.  
The fine binning and unprecedented quantity of the 
data impose tight constraints on PWA, especially 
at high$-L$ multipoles and at high CM energies 
where new resonances are expected to exist.

The three light quarks can be arranged in 6 baryonic 
families, the $N^\ast$, $\Delta^\ast$, $\Lambda^
\ast$, $\Sigma^\ast$, $\Xi^\ast$, and $\Omega^\ast$.  
The number of family members that can exists is not 
arbitrary.  Rather, the following proportionality is 
expected when the $SU(3)$-flavor symmetry of QCD is 
controlling symmetry~\cite{Ben}
\begin{equation}
        2~N^\ast : 1~\Delta^\ast : 3~\Lambda^\ast : 
	3~\Sigma^\ast : 3~\Xi^\ast : 1~\Omega^\ast .
\label{eq:eq5}
\end{equation}
Constituent quark models predict the existence of
no less than $64~N^\ast$ and $22~\Delta^\ast$ states 
with mass $< 3~GeV^2$~\cite{CaWi}.  Based on flavor 
$SU(3)$ symmetry, we expect to have twice as many 
$N^\ast$~($I = 1/2$) and $\Delta^\ast$~($I = 3/2$) 
resonances.  The number of experimentally identified 
resonances of each non-strange baryon family is 
$26~N^\ast$ and $22~\Delta^\ast$~\cite{PDG}.  The 
seriousness of the ``missing-states" problem is 
obvious from these numbers. The hypothesis of a 
very small $\pi N$ coupling of missing states has 
received support from a quark-model calculations
\cite{CaWi}.  We should stress that the standard 
$\pi N$ PWA (most of our current knowledge about 
the bound states of three light quarks~\cite{PDG}) 
reveals resonances with widths of order $\Gamma\sim$ 
100~MeV, but not too wide ($\Gamma > 500~MeV$) or 
possessing too small a branching ratio ($BR < 4\%$), 
tending (by construction) to miss narrow resonances 
with $\Gamma < 30~MeV$~\cite{Ar09}. However, 
conclusions on missing states should await the 
results of more realistic, coupled-channel 
calculations in which rescattering of the mesons, 
$\eta$, $\eta '$, $\omega$, and $\rho$, is considered.

\section{ACKNOWLEDGMENTS}

This material is based upon work at GW which supported 
by the U.S. Department of Energy, Office of Science, 
Office of Nuclear Physics, under Award Number 
DE-FG02-99-ER41110.


\end{document}